\documentclass[a4paper,11pt]{article}
\usepackage{jinstpub} 

\title{\boldmath Searching for WIMPs with TREX-DM: achievements and challenges}

\author{Juan~F.~Castel,}
\author{Susana~Cebri\'an,}
\author{Theopisti~Dafni,}
\author{David~D\'iez-Ib\'añez,}
\author{\'Alvaro~Ezquerro,}
\author{Javier~Gal\'an,}
\author{Juan~Antonio~Garc\'ia,}
\author{Igor~G.~Irastorza,}
\author{Mar\'ia~Jim\'enez,}
\author{Gloria~Luz\'on,}
\author{Cristina~Margalejo,}
\author{\'Angel~de~Mira,}
\author[1]{Hector~Mirallas\note{Corresponding author.}}
\author{, Luis~Obis,}
\author{Alfonso~Ortiz de Sol\'orzano,}
\author{Oscar~P\'erez,}
\author{Jaime~Ruz}
\author{and Julia~Vogel.}

\affiliation{Centro de Astropartículas y Física de Altas Energías (CAPA), Universidad de Zaragoza, Spain}

\emailAdd{mirallas@unizar.es}

\abstract{The \mbox{TREX-DM} detector, a low background chamber with microbulk Micromegas readout, was commissioned in the underground laboratory of Canfranc (LSC) in 2018. Since then, data taking campaigns have been carried out with Argon and Neon mixtures, at different pressures from 1 to 4 bar. By achieving a low energy threshold of 1~keV$_{ee}$ and a background level of 80~counts~keV$^{-1}$~Kg$^{-1}$~day$^{-1}$ in the region from 1~to~7~keV$_{ee}$, the experiment demonstrates its potential to search for low-mass WIMPs. Two of the most important  challenges currently faced are the reduction of both, background level and energy threshold. With respect to the energy threshold, recently a new readout plane is being developed, based on the combination of Micromegas and GEM technologies, aiming to have a pre-amplification stage that would permit very low energy thresholds, close to the single-electron ionization energy. With respect to the background reduction, apart from studies to identify and minimize contamination population, a high sensitivity alpha detector is being developed in order to allow a proper material selection for the \mbox{TREX-DM} detector components. Both challenges, together with the optimization of the gas mixture used as target for the WIMP detection, will take \mbox{TREX-DM} to explore regions of WIMP's mass below 1~GeV~c$^{-2}$.}


\begin{document}
\maketitle
\flushbottom

\section{Introduction}
\label{sec:intro}

The search for low mass WIMPs, which could be pervading the galactic dark halo, requires the use of light elements as target, detectors with very low energy threshold and very low background. Due to the lack of signal so far, it is interesting to extend the search to lighter WIMPs. Gas Time Projection Chambers (TPCs) with Micromegas planes have excellent features to fulfill these requirements and they have already demonstrated their good performance~\cite{CAST-Nature2017}.

\mbox{TREX-DM} (TPC for Rare Event eXperiments - Dark Matter) ~\cite{Iguaz2016} is a Micromegas-read High Pressure TPC of 20~L of active volume filled with Ar~(0.30~Kg) or Ne~(0.16~Kg) mixtures at 10~bar of pressure, for low mass WIMP search not focused on directionality. The detector was operated at surface in the University of Zaragoza as proof of concept and was approved by the Canfranc Underground Laboratory (LSC) in Spain, where it was installed at the end of 2018.

A brief summary of the challenges achieved by the experiment, from its installation at the LSC, is presented in section \ref{sec:Status}. After the good performance achieved during this first period, an update of the sensitivity prospects is presented in section \ref{sec:sensitivity}, where the main challenges of the project in the short-mid term are treated.

\section{Status of the experiment}
\label{sec:Status}

The installation of \mbox{TREX-DM} at the LSC was carried out at the end of 2018 with a mixture of atmospheric Argon~+~1\%~Isobutane at 1.5~bar~\cite{TPC_Paris2019}. At the beginning of 2019 the mixture was changed to Neon~+~2\%~Isobutane in order to avoid the background contribution of $^{39}$Ar and then the pressure was increased up to 4~bar.

The background model~\cite{Castel2019}, following a complete material screening program, pointed to expected levels of the order of 1-10~dru~(counts~keV$^{-1}$~Kg$^{-1}$~day$^{-1}$) in the region of interest from 0.2~to~7~keV$_{ee}$. The operation of the detector during 2019, with a gas mixture of Neon~+~2\%~Isobutane at 4 bar, allowed to quantify the background level in the low energy (LE) region from 1~to~7~keV$_{ee}$, being 2 orders of magnitude~(1000~dru) higher than that predicted by the background model. 

This background was dominated by Radon contamination in the gas system, mainly introduced by the moisture and oxygen filters. Some attempts to build simple radon cleanup filters were carried out, getting promising progress \cite{Altenmuller2021}. Finally, the Radon contamination was reduced by operating the system in open loop, without passing through the purification loop, with a very low flow ($\sim$~1~l/h). Once the mixture goes out the chamber it is not recovered but evacuated to the atmosphere. This type of operation in open loop caused the LE background reduction from 1000 to~80~dru~(1-7~keV$_{ee}$). 

The reduction of Radon in the gas system, thanks to the operation in open loop, allowed to discover alpha surface activity, previously masked by the volumetric Radon contamination. This surface activity, currently in the detector, is likely due to the Radon progeny attached to the surfaces exposed to the gas mixture inside the chamber. Some actions are planned to eliminate it (see section~\ref{sec:BackgroundReduction}). The current LE background level in \mbox{TREX-DM}~(80~dru) is likely dominated by these HE events from the nearest surfaces to the active volume. 

On the other hand, during the 2019-2022 data taking campaigns, the lowest energy threshold reached by the experiment was 1-1.5~keV$_{ee}$, depending strongly on the noise conditions and the gain achieved in the Micromegas, being even less than 1~keV$_{ee}$ during the last data taking campaign in 2022. The first sensitivity prospects of the experiment \cite{Iguaz2016} were estimated under the assumption of an energy threshold of 0.4~keV$_{ee}$ and, although the current level is close to the one predicted, some actions are planned in order to reduce this level down to 50~eV$_{ee}$ (see section~\ref{sec:EnergyThreshold}).

In October 2022, when the parts with surface contamination were going to be replaced, the experiment was completely dismounted after the request of the LSC management and stored in a safe place, waiting to be re-installed in a new site. In spring 2023, the experiment was re-installed in the Lab2500 at LSC and the detector has been taking data, from July 2023, to verify that all the systems work properly and verify that the background remains at the same level as in the previous site. Both points were verified. Once a clean area is installed by the LSC in the new site, as part as its services to the experiment, the chamber will be able to be opened and the works towards the reduction of the alpha surface contamination will be carried out, together other tasks towards the improvement of the performance of the detector.

\section{Sensitivity improvement to low-mass WIMPs}
\label{sec:sensitivity}

\mbox{TREX-DM} was built to search for WIMPs with a mass of the order of 10~GeV~c$^{-2}$ or less, using Neon as a main gas or even Argon depleted in $^{39}$Ar. Due to the lack of signal in the search so far, it is interesting to extend the search to lighter WIMPs, below~1~GeV~c$^{-2}$. In order to have the opportunity of reaching these regions of WIMP's mass, the current levels of background~(80~dru) and energy threshold~(<~1~keV$_{ee}$), achieved in \mbox{TREX-DM} in 2022, must be improved. In addition, the increase of Isobutane in the mixtures lead to higher sensitivity to lower WIMP's masses, below~1~GeV~c$^{-2}$, mainly due to the lower nuclei mass of the target.

\begin{figure}[htbp]
\centering
\includegraphics[width=1.0\textwidth]{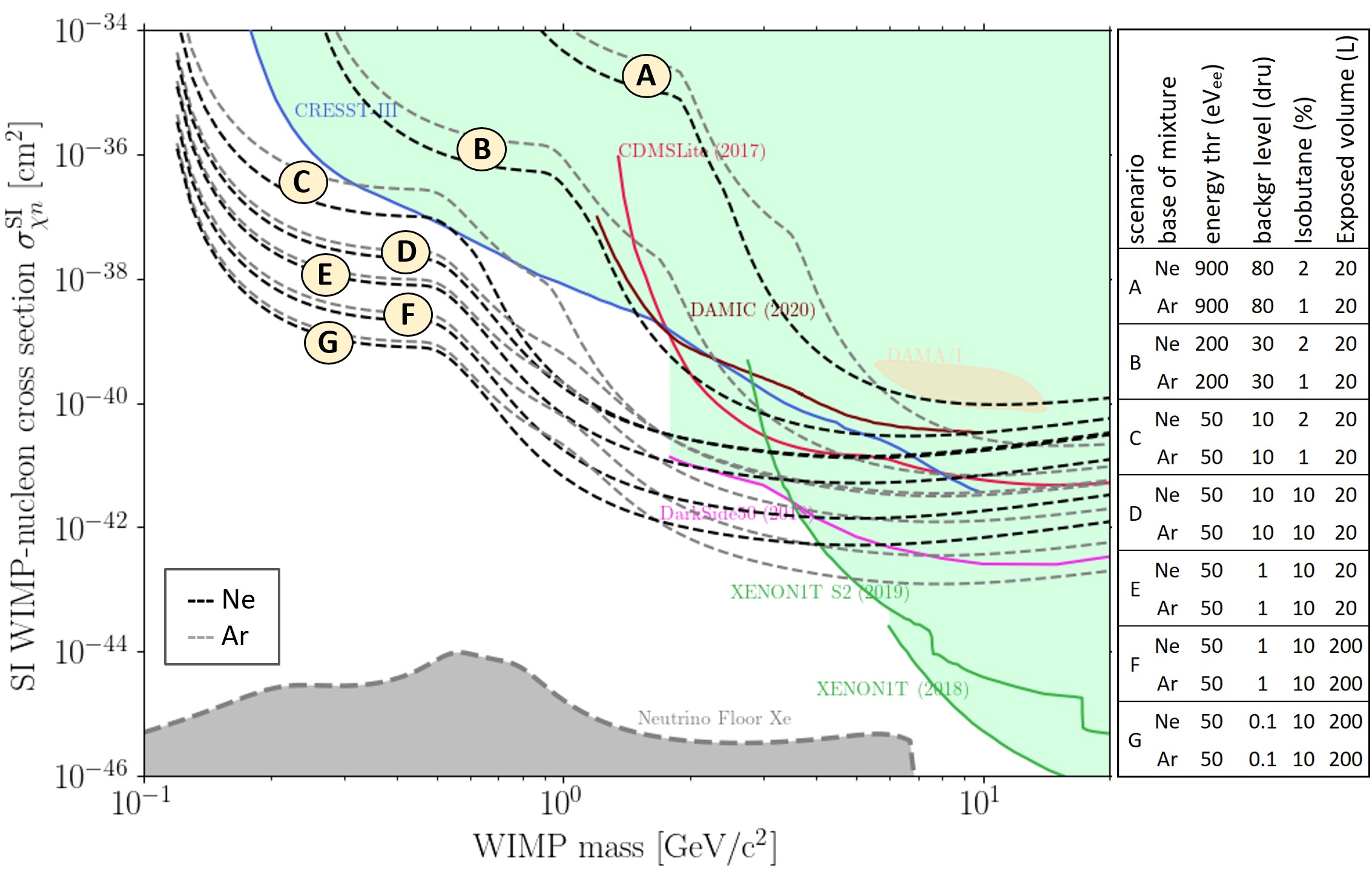}
\caption{WIMP-nucleon cross-section vs. WIMP mass, with current bounds from experiments and \mbox{TREX-DM}, under different conditions for \mbox{TREX-DM} shown in the table at the top-right and 1 year of exposure time. Each scenario is plotted with \mbox{Ne-based}~(black) and \mbox{Ar-based}~(grey) mixtures. \label{fig:WIMP_ExclussionPlot}}
\end{figure}

Figure~\ref{fig:WIMP_ExclussionPlot} shows several sensitivity curves of the experiment for 1 year of exposure time, assuming spin independent interaction, standard values of the WIMP halo model, astrophysical parameters and detector's parameters as energy threshold, background level or quantity of Isobutane. The plotted curves represent the scenarios that \mbox{TREX-DM} will go through as the named detector's parameters are gradually improved: scenario A achieved during the past campaigns, scenarios \mbox{B-E} planned for short-mid term and scenarios \mbox{F-G} for long term with a factor~10 scaled up chamber. 

\subsection{Background reduction}
\label{sec:BackgroundReduction}

The components that enter in direct contact with the active volume of the detector, and therefore would possibly contribute to the measured background levels, are the Micromegas planes, the field cage, the cathode and the gas itself. The microbulk Micromegas installed in \mbox{TREX-DM} in 2018 for physics runs at the LSC~\cite{Aznar2020} were replaced by new microbulk Micromegas in 2022. The new design is more robust from the point of view of the operation stability but also more radiopure, since a novel clean process to reduce the $^{40}$K during the fabrication has been applied, reducing the contamination of $^{40}$K a factor 3 respect to the previous version measured in~\cite{Castel2019}. These measurements estimate the contribution of the Micromegas to the background level of the order of 1~dru.

A comprehensive study of the background origin has led to the conclusion that the main contribution is due to alpha particles surface contamination, mainly coming from the cathode, likely due to the Radon progeny attached to it. The cathode consists of a copper frame where a thin foil of aluminized Mylar is tensed. Changing this element with a new one is one of the first steps to take as soon as a clean area is installed in the new site.

A fiducial cut of 18~x~18~cm$^2$ out of the total 25~x~25~cm$^2$ of the Micromegas has allowed to quantify the HE events emitted only from the cathode, but there is still an important population of events emitted from the field cage that, as in the case of the cathode, are likely due to the Radon progeny. These events can be discriminated by topology studies at the expense of the detector sensitive volume. Once the chamber will be able to be opened, the walls of the field cage, made of kapton-copper circuits, will be replaced too.

An exhaustive material radioassay campaign for \mbox{TREX-DM} has been carried out~\cite{Iguaz2016,Castel2019}, mainly based on germanium gamma-ray spectrometry but complemented by other techniques like GDMS or ICPMS. However, these techniques are not adequate to measure concentrations of particular isotopes that can produce alpha surface contamination, like $^{210}$Pb or $^{210}$Po. For this purpose, a new Micromegas-based camera for high-sensitivity screening of alpha surface contamination is being developed, the so called AlphaCAMM (Alpha CAMera Micromegas)~\cite{AlphaCAMM_Altenmuller2022}. The materials measurements planned to be done with this detector will allow to select the proper raw materials to be installed inside the \mbox{TREX-DM} chamber, reduce the background from HE events and therefore the LE background in the region of interest.

\subsection{Energy Threshold reduction}
\label{sec:EnergyThreshold}

As mentioned before, for exploring the low-mass WIMP parameter space, is fundamental to have the minimum energy threshold level. Figure \ref{fig:WIMP_ExclussionPlot} shows that the reduction from 900~eV$_{ee}$ (minimum energy threshold achieved in \mbox{TREX-DM} in 2022) to 50~eV$_{ee}$ has a very important impact in the sensitivity in the low-mass regions below~1~GeV~c$^{-2}$. 

\begin{figure}[htbp]
\centering
\includegraphics[width=1.00\textwidth]{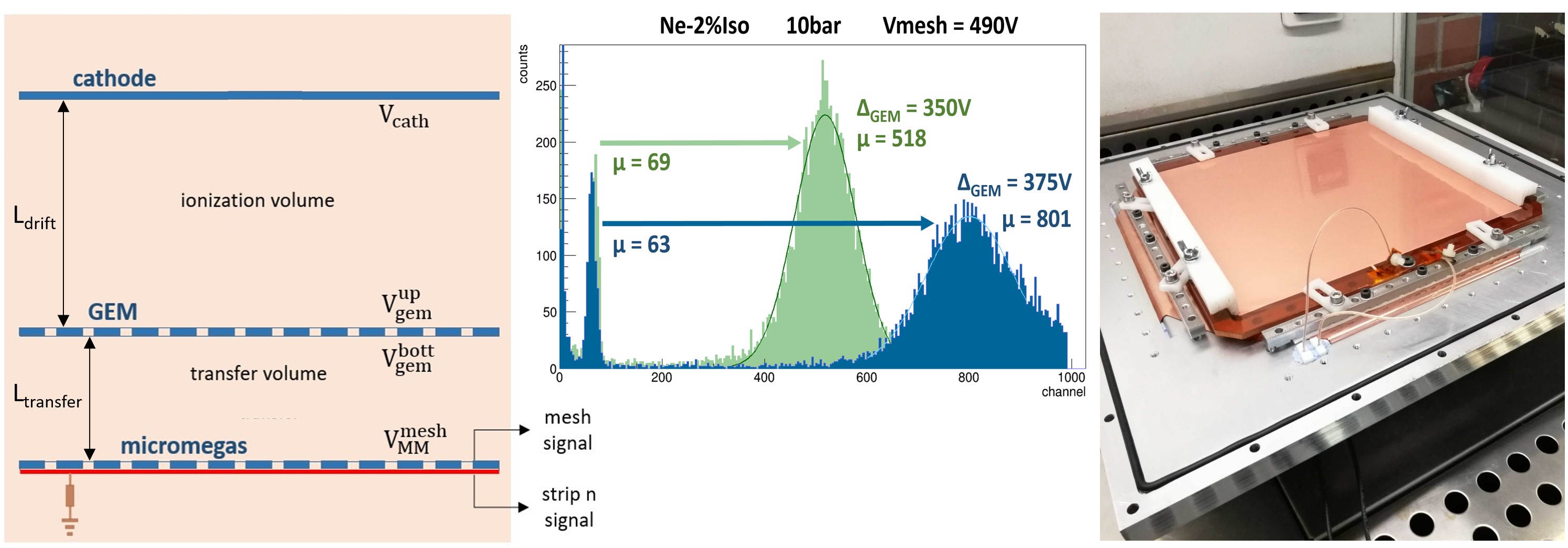}
\caption{\textbf{Left}: GEM-Micromegas system scheme. The primary electrons produced by a particle crossing through the ionization volume drift toward the GEM, where a first electron multiplication stage is done. The resulting electrons from this first multiplication go into the transfer volume and drift towards the Micromegas, where the second and final electron multiplication stage is done. \textbf{Center}: energy spectrum obtained with a simplified setup consisting of 2~cm~diameter GEM~+~Micromegas, in a 2~liters chamber, with Ne~+~2\%~Isobutane at 10~bar. The peaks with ($\mu$=518 and $\mu$=801) and without ($\mu$=69 and $\mu$=63) pre-amplification stage can be observed. \textbf{Right}: 25~x~25~cm GEM installed on top of the new version of the \mbox{TREX-DM} Micromegas, during the tests carried out in the laboratory of the University of Zaragoza. \label{fig:GEM-MM}}
\end{figure}

The energy threshold depends strongly on the gain achieved in the Micromegas, which depends in turn on the electric field applied in the amplification gap~\cite{Giomataris1996}. In order to increase the gain a new development has been carried out, based in the introduction of a new electron pre-amplification stage by means the installation of a GEM~(Gas Electron Multiplier) on top of the Micromegas~\cite{KANE2002}. Data were taken in the laboratory (figure \ref{fig:GEM-MM}) using Ar~+~1\%~Isobutane and Ne~+~2\%~Isobutane from 1 to 10~bar, achieving pre-amplification factors from 12 (Ne~+~2\%~Isobutane at 10~bar) to 100 (Ar~+~1\%~Isobutane at 1~bar). Pre-amplification factors depend on the pressure of the gas, but even in the worst scenario (factor 12 with Ne~+~2\%~Isobutane at 10~bar) the system GEM-Micromegas could potentially reach energy thresholds down to single-ionization electron. 

The GEM foils are ready to be installed in \mbox{TREX-DM} (figure \ref{fig:GEM-MM} right) and, once the chamber can be opened, the GEM will be installed together with the new cathode and the new field cage. 

\subsection{Gas mixture optimization}
\label{sec:Isobutane}

The first sensitivity prospects of \mbox{TREX-DM}~\cite{Iguaz2016} were made for a quantity of Isobutane such that the mixture could be classified as non-flammable (1\% in Argon and 2\% in Neon). These prospects was better for Ne than for Ar-based mixtures (scenarios \mbox{A-C} in figure \ref{fig:WIMP_ExclussionPlot}), but increasing of Isobutane reduces the difference between the two mixtures, as noted when comparing both mixtures in scenarios \mbox{D-G}. And most importantly, the increase of Isobutane in the mixtures lead to higher sensitivity to WIMP's masses below~1~GeV~c$^{-2}$, mainly due to the lower nuclei mass of the target. 

Given the evidence of improvement due to Isobutane, and taking advantage of the more isolated new site where \mbox{TREX-DM} is installed, the LSC recently accepted the request to operate with flammable mixtures up to 10\% of Isobutane. This quantity of Isobutane is a \mbox{trade-off} solution that improves much the sensitivity but with an acceptable risk, taking into account that the improvement of the sensitivity with mixtures over 10\% is not so notable as when passing from 1 or 2 to 10\%.

The use of the new flammable mixtures forces the application of new safety measurements in the site of the experiment. These safety measurements are being currently designed and will be implemented in the \mbox{short-mid} term.

\section{Conclusions}
\label{sec:Conclusions}

\mbox{TREX-DM} has demonstrated its good performance during data taking campaigns carried out from its installation underground. Although the current background level is almost one order of magnitude higher than that predicted by the background model, the origin of great part of this background has been identified and short-term actions will be taken in order to reduce it. Also, the energy threshold achieved during data taking campaigns is close to the first prospects, but the new prospects are more ambitious and a new combination GEM-Micromegas has been developed to take the detector close to the single-electron detection.

The good performance, together the future plans to reduce the background, the energy threshold and optimize the gas mixture, demonstrates the \mbox{TREX-DM} potential to explore unexplored regions of WIMP's mass below 1~GeV~c$^{-2}$.


\acknowledgments

We acknowledge support from grant PID2019-108122GB-C31 funded by MCIN/AEI/10.13039/ 501100011033, from "European Union Next Generation EU/PRTR" (Planes complementarios, Programa de Astrof\'isica y F\'isica de Altas Energías), from the European Union's Horizon~2020 research and innovation programme under the European Research Council (ERC) grant agreement ERC-2017-AdG788781 (IAXO+). We acknowledge the technical support from LSC and GIFNA~staff. Also, we would like to acknowledge the use of SAI~(Servicio General de Apoyo a la Investigaci\'on) from Universidad de Zaragoza.


\bibliographystyle{JHEP}
\bibliography{biblio.bib}



\end{document}